\documentclass[aps,amssymb,amsmath,prb,reprint,noshowpacs,]{revtex4-1}

\usepackage{times}
 
\usepackage{graphicx}% Include figure files
\usepackage{bm}% bold math
\usepackage{epsfig}
\usepackage{scalefnt}

\usepackage[usenames]{color}
 
\newcommand{\narrowfigwidth}{8.6cm}%% columnwidth 
\newcommand{\widefigwidth}{17.2cm}%% twice columnwidth

\begin{document}
%\scalefont{1.05}
\title{Lasing at the band edges of plasmonic lattices}% Force line breaks with \\
 
\author{A. Hinke Schokker}
\affiliation{Center for Nanophotonics, FOM Institute AMOLF, Science Park 104, 1098 XG Amsterdam, The Netherlands}
\author{ A. Femius Koenderink}
\affiliation{Center for Nanophotonics, FOM Institute AMOLF, Science Park 104, 1098 XG Amsterdam, The Netherlands}
\email{f.koenderink@amolf.nl}
%\url{www.amolf.nl}

\date{\today}% It is always \today, today,

\begin{abstract}
We report room temperature lasing in two-dimensional diffractive lattices of silver and gold plasmon particle arrays embedded in a dye-doped polymer that acts both as waveguide and gain medium. As compared to conventional dielectric distributed feedback lasers, a central question is how the underlying band structure from which lasing emerges is modified by both the much stronger scattering and the disadvantageous loss of metal. We use spectrally resolved back-focal plane imaging to measure the wavelength- and angle dependence of emission below and above threshold, thereby mapping the band structure. We find that for silver particles, the band structure is strongly modified compared to dielectric reference DFB lasers, since the strong scattering gives large stop gaps. In contrast, gold particles scatter weakly and absorb strongly, so that thresholds are higher, but the band structure is not strongly modified. The experimental findings are supported by finite element and fourier modal method calculations of the single particle scattering strength and lattice extinction.

\end{abstract}

\pacs{Valid PACS appear here}
\maketitle

In the past decade, plasmonics has become a very active field of research within optics owing to the unique opportunities for broadband strongly enhanced light matter interaction in precisely fabricated metal nanostructures\cite{maier2007}. Enhanced light matter interaction arises from the fact that  plasmons, as hybrids of photons and charge density oscillations, are not restricted to the conventional diffraction limit.  In addition, it has been shown that plasmon particles can enhance emission decay rates of fluorophores due to high Purcell factors over large bandwidths \cite{agio2013,anger2006,kuehn2006,mertens2007,giannini2011}. In fact, for low quantum efficiency fluorphores up to 1000-fold brightness enhancements per molecule have been reported near bow tie antennas \cite{punj2013,kinkhabwala2009,anger2006,kuehn2006,mertens2007,giannini2011}.  Huge field enhancements (of order 10$^3$ in electric field) have further been evidenced in surface enhanced Raman and surface enhanced infrared spectroscopy \cite{stiles2008,le2008}. 
 Interest in exploiting plasmonics for lasing was sparked by the seminal paper by Bergman and Stockman in 2003 \cite{bergman2003}, where plasmonics was proposed for reaching deeply sub-diffraction sized lasers, ultralow thresholds, ultrafast laser dynamics, and unique properties due to the fact that only a few gain molecules and intracavity photons participate\cite{stockman2010}.  This vision of a `spaser' where lasing occurs due to nanoscale amplification of dark plasmons has led to a suite of recent experiments focusing on the smallest plasmonic lasers, targeting colloidal metal particles with gain\cite{noginov2009}  as well as hybrid plasmon modes confined in a narrow gap between a metal film and II-VI or III-V nanowires that provided the gain \cite{oulton2009,ma2011,kwon2010,lu2012,hill2007,Khajavikhan2012}.

 Aside from efforts to realize the highest possible field enhancements in narrow gaps of single structures, many efforts in plasmonics have been devoted to light-matter interaction in oligomers of scatterers and periodic lattices. Indeed Yagi-Uda phased array antennas\cite{li2007,koenderink2009,kosako2010,curto2010,arango2012}, Fano resonant oligomers\cite{zhang2008,lukyanchuk2010,verellen2009,hentschel2010,lassiter2010} and periodic lattices\cite{lozano2013,murai2013,rodriguez2012,vecchi2009,haynes2003,zou2004,abajo2007} are among the most practical structures not only to control field enhancement but also to obtain a balanced trade off between enhancement,  Ohmic loss,  and directivity control for emitters. In particular, in diffractive lattices single particle plasmon resonances can hybridize with Rayleigh anomalies or with planar waveguide modes  to form extended collective modes\cite{waele2007}. These systems have been shown to be very practical for improving broad area emission devices such as LEDs and phosphors,  allowing simultaneous control over emission directivity and rate, at much lower losses than offered by single particle resonances\cite{lozano2013,murai2013,rodriguez2012,vecchi2009}.  The picture that has emerged is that plasmonic structures can on one hand significantly enhance emission brightness from intrinsically very inefficient  emitters by use of Purcell enhancement  as a means to help radiative decay to  outcompete nonradiative processes. On the other hand,  in realistic application scenarios for solid-state lighting, already very efficient emitters can not benefit from plasmonics through Purcell enhancement, but do benefit through plasmonic band structure effects that ensure redirection of light  into select angles. In this case, the most efficient redirection is obtained  through extended, not strongly localized, plasmon modes.

 In the context of lasing, diffractive plasmon lattices were first studied by Stehr et al., \cite{stehr2003} who reported a metallic particle grating based laser and showed linewidth narrowing and threshold behavior in these systems. A complementary geometry was reported very recently by van Beijnum et al. \cite{beijnum2013} who demonstrated a plasmon lattice laser based on hole arrays in gold paired to a III-V quantum well gain medium. Suh et al., and Zhou et al. \cite{suh2012,zhou2013}, finally, reported on lasing in bow tie and nanodisk arrays, i.e., in particle arrays similar to those reported by Stehr et al. \cite{stehr2003} 
As in the case of spontaneous emission enhancement, plasmonic effects can impact lasing through two effects.  On one hand,  Purcell enhancements and  near field enhancement can accelerate emission dynamics. On the other hand,  even in absence of strong Purcell enhancements the formation of a plasmonic band structure with large stop gaps could modify the 
distributed feedback mechanism. The work of Suh et al. and Zhou \emph{et al.}\cite{suh2012,zhou2013} focused particularly on the role of plasmonic Purcell enhancements in lasing,  for which reason an intrinsically very poor efficiency gain medium was chosen. Thereby, only the dye in very close proximity to the metal that experienced rate enhancement participated in the lasing.  Here we focus on the more application relevant scenario of plasmonic lasing in an efficient gain medium, in which case the main questions that arise  are how the band structure of plasmonic lattice lasers differs from that of conventional 2D distributed feedback (DFB) lasers due to the plasmonic nature of its constituents, and how the trade off between much larger scattering strength and disadvantageously large loss of metal particles influences the lasing behavior.

 In this paper we report a comprehensive lasing study on particle array lasers fabricated from square lattices of silver (Ag),  gold (Au)  and as non-plasmonic reference titanium dioxide (TiO$_2$) embedded in a dye-doped polymer that at the same time acts as gain medium and supports a waveguide mode. We aim at uncovering what the band structure and lasing conditions of such systems are as a function of scattering strength (highest for Ag particles) and loss (highest for Au particles). Therefore we use a high efficiency dye as would be used in prospective solid-state applications, and operate in a regime where the effects of Purcell enhancements are expected to be small. 
  To answer these questions we have implemented a new measurement technique to map below-threshold emission and lasing in energy-momentum diagrams that can be acquired in a single shot in our sub-nanosecond optically-pumped setup, and that span the entire angular collection range of a high-NA objective. We analyze  the plasmonic band structure and for Ag arrays find stop gap widths far in excess of those in dielectric DFB lasers and similar to those in the reported metal hole array laser of van Beijnum\cite{beijnum2013}. 
The paper is structured as follows.  In section \ref{section:experimental} we explain the setup, materials and measurement procedure. In section \ref{section:results} we present emission spectra,  measured 2D Fourier space distributions of emission and energy-momentum diagrams, all below and above threshold.  In section \ref{section:theory} we interpret our band structure measurements in terms of calculated single particle scattering properties obtained from finite element simulations,  and in terms of calculated angle-dependent extinction obtained with a rigorous coupled wave analysis (RCWA) method.

\section{Experiment\label{section:experimental}}
\begin{figure}
\centerline{\includegraphics[width=\narrowfigwidth]{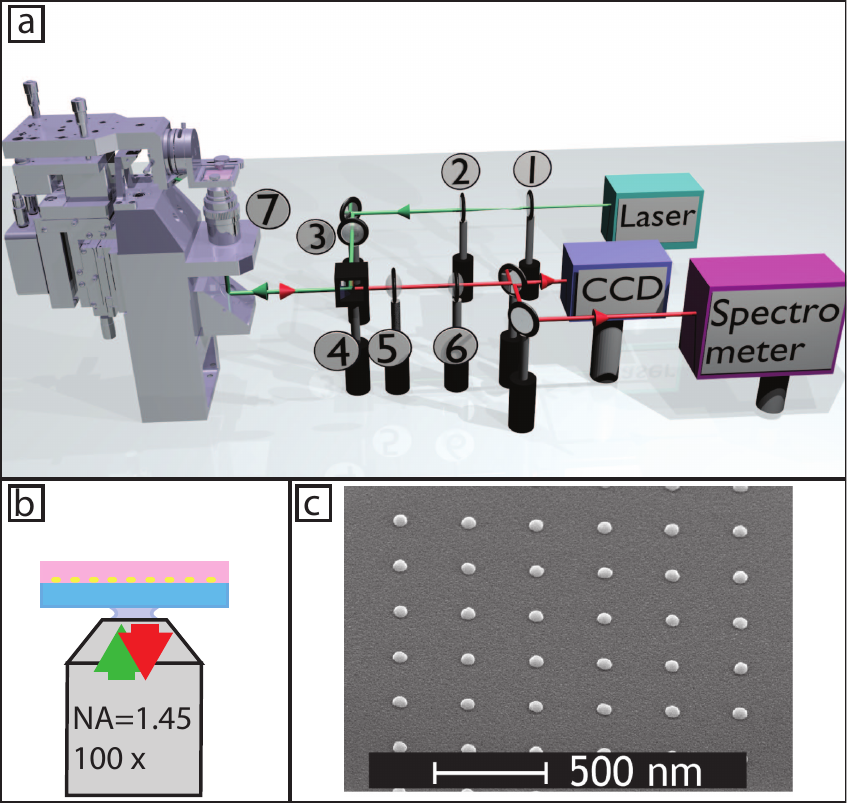}}
\caption{\label{fig:setup} (a) Schematic of the setup. We illuminate the sample with laser light ($\lambda$=532 nm) from the glass side. We measure the fluorescence using the CCD and spectrometer. Right after the laser there are two lenses (1 and 2) that constitute a telescope to increase the beam diameter. After reflection of a mirror, there is an epi-lens (3) followed by a filter cube (4) that contains the dichroic mirror and a long pass filter, allowing only fluorescent red light on the detector side. On the right side of the filter cube there is a flippable fourier lens (f=200 mm)(5) followed by a tube lens (f=200 mm)(6). The objective is mounted in a specially designed microscope mount (7). Not shown in the image are the AOM, together with a polarizer just in front of the telescope, to tune the laser power. In (b) we show a schematic of the objective with the sample. In (c) a scanning electron micrograph  of one of the fabricated particle arrays is shown.}
\end{figure}
\subsection{Sample fabrication}
We use Menzel glass cover slides of 24 x 24 x 0.17 mm that have been cleaned in  a solution of H$_2$O, H$_2$O$_2$ and NH$_4$OH at 75 $^\circ$C. After cleaning, we spincoat a positive resist to define our structures. For this we use the electron beam resist ZEP520 diluted in a ratio of 5:2 with anisole for which spincoating at 1500 rpm  results in a layer thickness of ~150 nm. With electron beam lithography we define hole arrays in a square pattern using dot exposures between 0.001 pC and 0.002 pC using an electron gun voltage of 20 kV and a current of 0.031 nA. We vary the lattice constant from 350 nm to 500 nm in steps of 10 nm. The hole size is ~100 nm. The hole arrays are 200 $\mu$m in size so that in optical experiments explained below, the arrays exceed the optical pump spot in diameter. To fabricate silver particle arrays we subsequently deposit 2 nm of chromium followed by 30 nm of silver by thermal evaporation, performed at a pressure of $< 10^{-6}$ at an evaporation rate of 0.5-1 \AA/s. For the titanium dioxide samples we directly deposit 30 nm of titanium dioxide using electron beam deposition. We perform lift-off by immersing the samples in N-methyl pyrrolidone (NMP) at 65$^\circ$ for 5 minutes. For silver, lift-off is achieved by leaving them overnight at 50$^\circ$ in anisole, as NMP degrades silver. After lift off the samples are rinsed in isopropanol and blow dried with nitrogen. Figure \ref{fig:setup}c  shows an Ag particle array resulting from the fabrication procedure. 

 To obtain  a waveguide with gain, we use the negative photoresist SU8 and dope it with rhodamine 6G by mixing 5.25 mg of Rh6G perchlorate with 1 mL of cyclopentanone (the solvent for SU8).  The cyclopentanone with Rh6G is added to 1 ml of SU8-2005, after which we ultrasonicate the solution for 10 minutes. The final solution has a Rh6G perchlorate concentration of 0.25 wt\%. We spincoat the SU8 solution on the particle array samples at 3000 rpm, resulting in a ~450 nm thick SU8 layer. This thickness results from a tradeoff between two requirements: on one hand  sufficiently small thickness to ensure single waveguide mode operation, and on the other hand sufficiently large thickness to ensure good mode overlap with the gain medium. 
  After spincoating, we bake the samples for 2 minutes at 95 $^\circ$C to evaporate the excess cyclopentanone. 
Prepared as such, the SU8 is not cross-linked, enabling removal of the SU8 layer with acetone after performing measurements on the samples.

\subsection{Experimental set-up\label{section:experimentalsetup}}
We use an inverted fluorescence microscope as shown in figure \ref{fig:setup}(a). In this setup, the sample is mounted with the glass side down (close up sketch panel (b)), and both pump and detection occur through the objective, i.e. from the glass side. We use a home-built microscope tower, the most important pieces of which are an objective (Nikon, Plan Apo $\lambda$ 100x /1.45 NA) fixed to the microscope frame and a sample mount that can be translated in XYZ relative to the objective using micromechanical and piezo controls. The sample is pumped using a 532 nm pulsed laser (Teem Photonics, type STG-03E-1S0) which has a pulse width of 500 ps and a maximum energy per pulse of 4.5 $\mu$J. We use an epi-lens in the pump path, resulting in a parallel beam  with a diameter of 70 $\mu$m  emerging from the objective. The laser power is computer controlled by an acousto-optical modulator (AOM). We monitor the resulting pump power in real time with a home built pulse integrator. To filter out unwanted reflected pump light the fluorescence is sent through a long pass filter (Chroma, HHQ545lp)  after passing the dichroic mirror (Semrock, Di01-R532-25x36). Fluorescence  is detected by either a thermoelectrically cooled (Andor CLARA) Si CCD camera or a Shamrock303i spectrometer with an (Andor Ivac) Si CCD detector. To focus the light on the CCD and spectrometer entrance slit, we use an f=200 mm tube lens. The pump laser can fire single pulses allowing single shot measurements when triggering the laser, CCD and spectrometer simultaneously. Single shot exposure minimizes sample damage caused by bleaching of the Rh6G when performing a sequence of measurements for varying pump power. In addition to collecting images and spectra in real space, we do fourier imaging by adding a lens on a flip mount at a focal distance from the back focal plane of the objective \cite{sersic2011,lieb2004,thomas2007,randhawa2010,alaverdyan2009,huang2008}.

  Fourier imaging maps the back focal plane of the objective onto the CCD camera, providing direct information on angular emission.  The high NA objective (NA=1.45) allows for a large maximum collection angle of $\theta=73 ^\circ$ in glass, enabling us to  image a large part of $k_{||}$ space. We note that the 2D back focal plane images we collect in this fashion on the Clara CCD camera are panchromatic images, i.e., not separated in spectral components. Ideally one would measure a spectrally resolved Fourier image, since  a spectrally resolved fourier image would be a direct map of the dispersion diagram. This can be done by scanning a fiber which is coupled to a spectrometer through the entire fourier image, or by imaging a slice of the fourier image centered at $k_x$=0 onto the slit of an imaging spectrometer \cite{taminiau2012}.  As we aim at single-shot measurements the latter method is preferred.
The Andor IVAC camera contains a CCD chip with 200 x 1650 pixels. To make spectral fourier images we set the spectrometer imaging mode to full imaging resulting in a full spectrum for 200 points along the $k_y$ axis.

 For every particle array we start by taking fourier images of the fluorescence of single pump pulses as a function of input power by increasing the AOM voltage linearly in 200 steps from 0\% to 50\% of its maximum value. Subsequently, we flip in the mirror, sending the light to the spectrometer. We center the fourier image on the spectrometer slit by fully opening the slit, observing the image in 0$^{th}$ order and moving the fourier lens transversally until the circular fourier image is in the center of the image of the slit. For fourier spectra, we need to add the fluorescence resulting from 50 pulses because the light is spread over a large detector area. We find that 50 pump pulses do bleach the sample noticeably for higher pump powers. To make sure we see clear signs of lasing before the sample has bleached, we start with high pump powers (above lasing threshold) at an AOM percentage of 50\% and go down in 200 steps to 0\%. From the fourier spectra we calculate real space spectra by integrating over $k_y$ for each wavelength. Because of bleaching the threshold pump powers deduced from the fourier images are slightly lower than the threshold we find from the fourier spectra.

\section{Results\label{section:results}}
\begin{figure}
\centerline{\includegraphics[width=\narrowfigwidth]{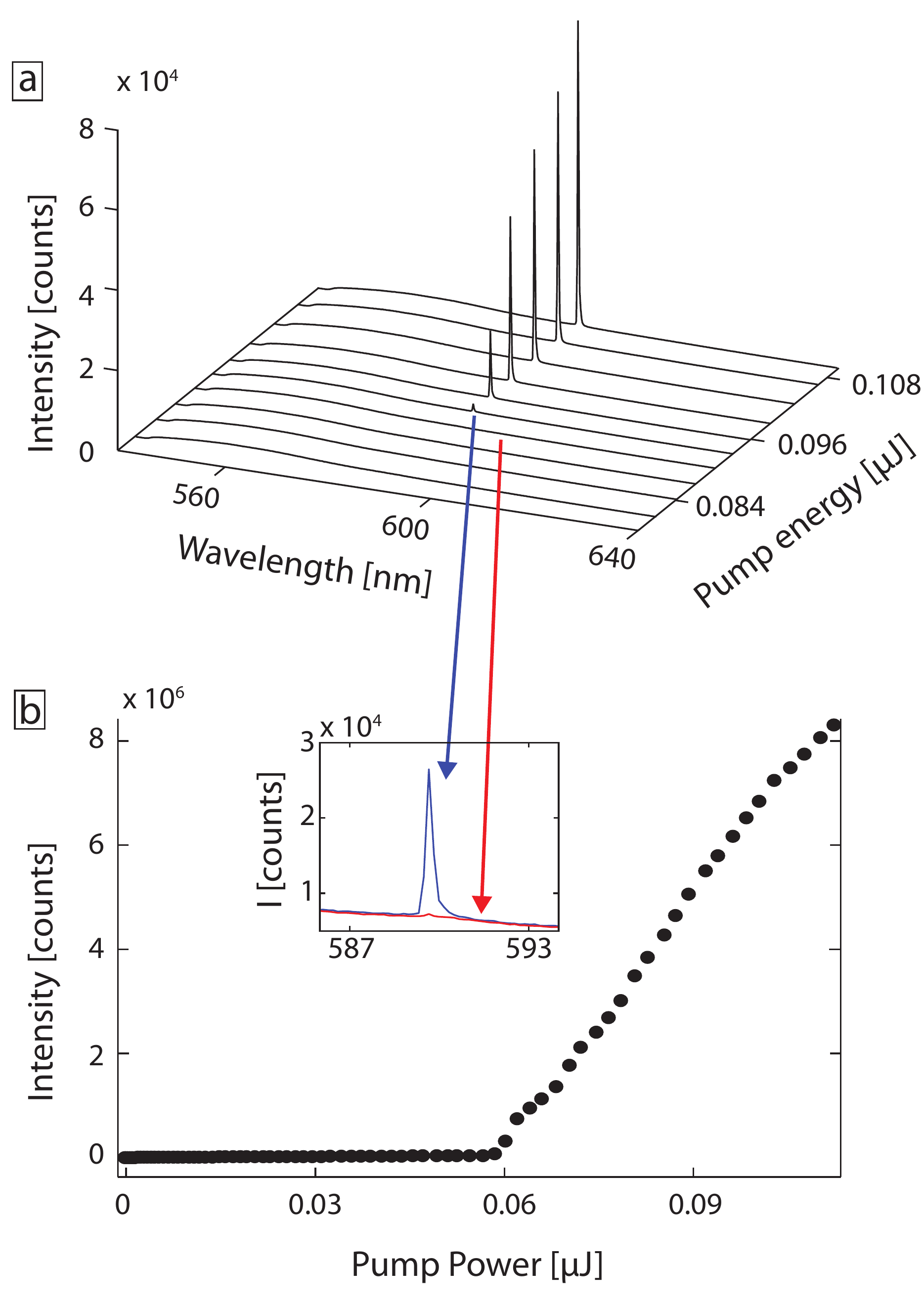}}% Here is how to import EPS art
\caption{\label{fig:wf}(a) Plot of emission spectra for different  pump powers measured by imaging an Ag particle array (pitch 380 nm, particle diameter 100 nm)   onto the spectrometer slit measured upon excitation with a single pump pulse. A clear threshold behaviour can be seen from the sharp peak occurring for pump powers above 59 nJ. Panel (b)  threshold curve, plotting the area under the lasing peak versus pump pulse enery. The inset shows two spectra just above and just below threshold.}
\end{figure}
\subsection{Spectra}
Figure \ref{fig:wf}(a) shows a waterfall plot of spectra for increasing pump pulse power for a silver particle array with a pitch of 380 nm and particle diameter of 100 nm. These spectra are obtained by applying full vertical binning over the central part of the fourier image, thus including all angles along the $k_y$ axis. At a pump pulse energy of 59 nJ (corresponding to a pulse irradiance of 1.53 mJ/cm$^2$)  a clear peak emerges at a wavelength of 589 nm which dominates the emission spectrum for all higher pump powers. This can be seen more clearly from the inset of figure \ref{fig:wf} where we plot a spectrum just below (red graph) and just above lasing threshold. The onset of the sharp peak is characteristic of lasing and the pump power at which it occurs is the lasing threshold. From the inset it can be seen that the lasing peak linewidth is on the order of a nanometer which is limited by the resolution of the spectrometer.

To construct a threshold curve, we define emission power as the total number of CCD counts under the lasing peak visible in figure \ref{fig:wf}(a), where we integrate over three spectral bins, corresponding to a total bandwidth of 0.5 nm. Figure \ref{fig:wf}(b) shows the emission power versus pump power. The lasing threshold can be recognized by a sharp kink at a pulse energy just below 60 nJ. The pulse energy density required to reach threshold is thus around 1.53 mJ/cm$^2$. This pulse energy density is comparable to that reported for plasmon particle arrays in a non-waveguiding polymeric gain layer by Zhou et al. \cite{zhou2013}, although that laser operated much further into the infrared. These thresholds are approximately 10 times above those typically required for purely polymeric DFB lasers, such as the 2D MEH PPV DFB laser reported by Turnbull \cite{turnbull2003}. Finally we note that Stehr et al. \cite{stehr2003}  reported thresholds about equal to those of Turnbull et al.\cite{turnbull2003}  for a gold particle array plasmon laser in a poly paraphenylene matrix.  Besides possible differences in gain coefficient, two possible explanations are the presence of ohmic damping in the plasmonic particles and the much stronger outcoupling through scattering that plasmon particles offer. 

\begin{figure} 
\includegraphics[width=8.6cm]{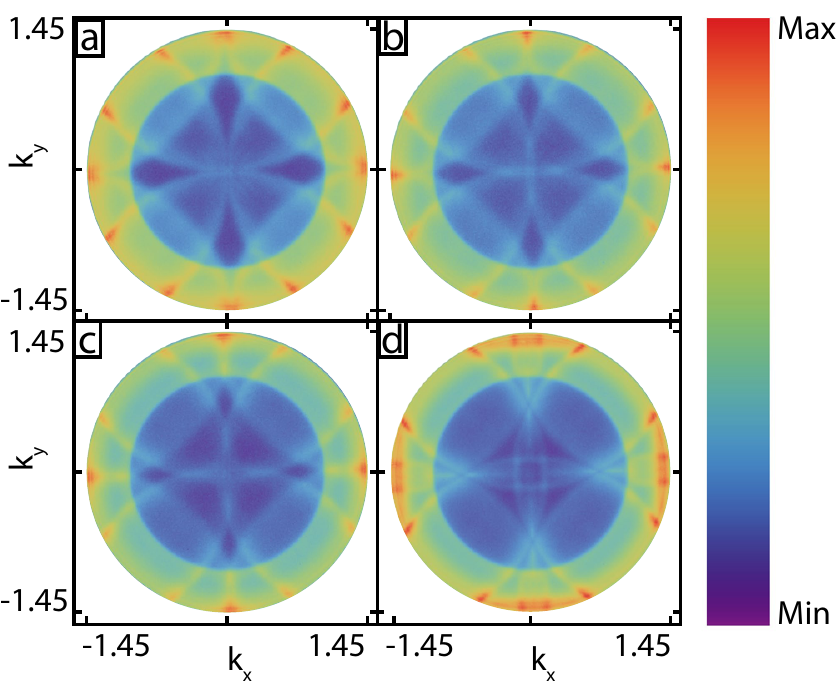}% Here is how to import EPS art
\caption{\label{fig:fourbelow}Fourier images below the lasing threshold for a pitch of a) 360 nm, b) 370 nm, c) 380 nm and d) 400 nm. The color map ranges are [223, 1196], resp. [102, 739], [128, 941] and [319, 1948]. The reported wave vector axes are normalized to $\omega/c$.}
\end{figure}
\begin{figure} 
\includegraphics[width=8.6cm]{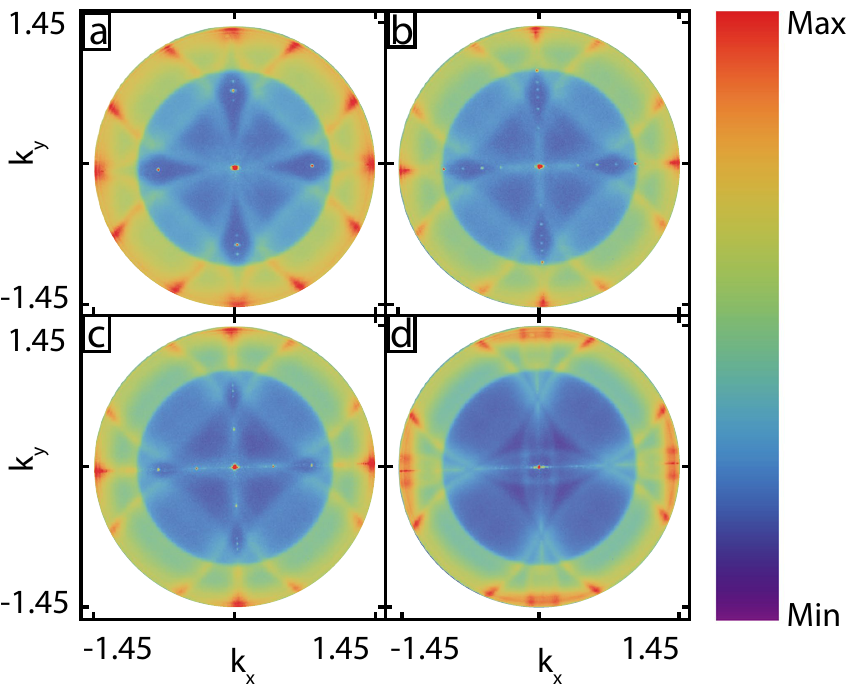}% Here is how to import EPS art
\caption{\label{fig:fourabove}Fourier images \emph{just} above the lasing threshold for a pitch of a) 360 nm, b) 370 nm, c) 380 nm and d) 400 nm. The color map ranges are the same as in figure \ref{fig:fourbelow}. Note the appearance of the narrow feature in the center of all images, which shows the onset of lasing emission. The reported wave vector axes are normalized to $\omega/c$.}
\end{figure}

In addition to measuring spectra as a function of pump power, we collect fourier images of the fluorescence as shown in figure \ref{fig:fourbelow} and \ref{fig:fourabove}, which report Fourier images just below and just above threshold respectively, for four particle lattices with pitches d=360, 370, 380 and 390 nm (panels a-d in both figures). In each figure, two features stand out independent of particle pitch. First, we see a high intensity ring where most of the emission exists, indicating that most below-threshold emission exits at large angles. The inner edge of this ring corresponds to an NA of 1 or equivalently, to the critical angle of the glass-air interface. The outer edge is set by the NA of the immersion oil objective. That fluorescence is preferentially emitted at angles just above an NA=1 is a well known feature  for  emitters on a glass-air interface \cite{taminiau2012,lieb2004,novotny2007}, and consistent with radiation pattern calculations according to chapter 10 of Ref.~\onlinecite{novotny2007}, which  show radiation patterns peaking   at the critical angle of the glass-air interface. 

 Second, we see higher intensity circles, displaced from the center and repeating in the $k_x$ and $k_y$ direction with a fixed period that changes with particle pitch. Indeed based on the wave vector scale calibration of our images we can confirm that the Fourier space periodicity corresponds to a square lattice with $\frac{2\pi}{d}$ period, i.e., to the reciprocal lattice of our structure. Based on the absolute wave vector scale calibration of our images, we can also convert the radius of the circles into a propagation constant. We find  $k_{\mathrm{circle}}/(\omega/c) =1.52\pm0.03$ , where the factor 1.52 corresponds very well with the calculated mode index for the fundamental TE and TM guided mode of the SU8 layer as calculated from Eq. 4.4 and 4.17 in Ref. \onlinecite{Urbach}. 

\begin{figure*}
\centerline{\includegraphics[width=\widefigwidth]{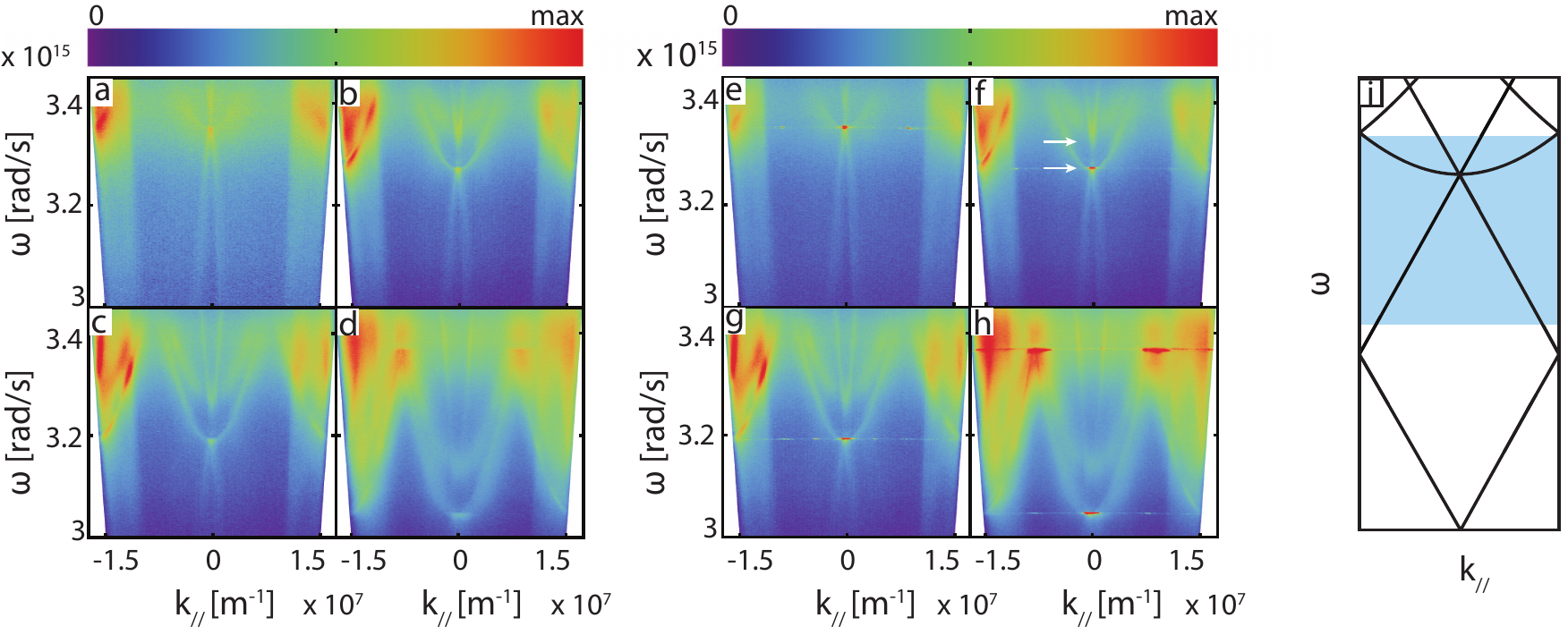}}
\caption{\label{fig:fourspecsbanddiagram} Fourier spectra for 4 different pitches, just below threshold (a-d) and just above threshold (e-h), and a schematic of a general banddiagram (i). Measurements are for a pitch of 360 nm (a and e), 370 nm (b and f), 380 nm (c and g) and 400 nm (d and h). The maximum value of the color bar are set at 230, 200, 200 and 250 counts respectively, in (a-d). Above threshold the maximum value of the colorbar is 230, 250, 200 and 300 counts, respectively (panels e-h). All colorbars start at 0. Note the lasing emission that appears as a narrow feature at $k_{||}=0$ (The horizontal lines across the diagram, surrounding the lasing peak, are CCD blooming artifacts). For clarity we have indicated the lower and upper stop band edge for second order diffraction by white arrows in panel (f). Lasing occurs at the lower edge. }
\end{figure*}
Overall the fluorescence Fourier pattern is a direct, single-shot CCD image of the repeated zone scheme iso-frequency surface of the waveguide mode dispersion that is well known to occur for periodically corrugated waveguides \cite{rigneault1999,harrison1970,kretschmann2002,langguth2013}. In other words, due to in-plane Bragg scattering that couples any $\mathbf{k}_{||}$ into $\mathbf{k}_{||}+\mathbf{G}$, the circular dispersion relation $\omega$, $\mathbf{k}_{||}$ of the waveguide mode of index n=1.52 repeats at every reciprocal lattice point $\mathbf{G}$. We expect that for each intersection of circles an anti-crossing should be visible, as the finite scattering strength of the plasmon particles should open up noticeable stop gaps in the nearly free photon dispersion approximation. However, any stop gaps that may occur at the crossing points are obscured in these images due to the fact that spectral averaging limits their sharpness. To overcome this problem we use the spectral imaging procedure described in the experimental section which gives us full dispersion diagrams of the emission over the entire  detectable $\mathbf{k}_{y}$ range. The resulting $\omega,k_y$ diagrams are shown in figure \ref{fig:fourspecsbanddiagram}(a-d,  just below threshold) and figure \ref{fig:fourspecsbanddiagram}(e-f, just above threshold). Again we can see high intensity bands corresponding to the high intensity ring in the fourier image. In addition, we distinguish a pair of steep straight lines that cross at the $\Gamma$ point ($k_y=0$) at a frequency of $3.4\cdot 10^{15}$ rad$\cdot$s$^{-1}$ for $d=360$~nm. Furthermore, a parabolic band with minimum at (or just above) the crossing of steep lines is evident, that has its minimum at or just above the crossing of steep lines. These features can be understood by looking at a generic dispersion diagram respresenting the folded free-photon dispersion, as indicated in figure \ref{fig:fourspecsbanddiagram}i. The straight lines that begin at the origin are the linear waveguide dispersion.

 At the first order Bragg condition $k_y=\frac{\pi}{d}$,  the free-photon dispersion copies shifted along $k_y$ by $\frac{2\pi}{d}$ fold back into the first Brillouin zone. At  twice this frequency the second order diffraction condition is met, as is evident from the fact that the folded dispersions again cross (straight lines). For a square lattice at the same frequency the diffraction condition is met from the grating vector $\mathbf{G}=(\frac{2\pi}{d},0)$ perpendicular to the $k_y$  axis.  This diffraction leads to the parabola. When the particle pitch increases, the first order Bragg condition is met at a lower frequency and all features move down as the waveguide mode circles repeat with a larger period. In figure \ref{fig:fourspecsbanddiagram}(e-h) we show fourier spectra just above the lasing threshold. Lasing spots are visible as high intensity spots that occur exactly at the crossing point of the lines with the parabola. This corresponds to the second order bragg diffraction condition. Qualitatively this behavior is exactly as generally observed for 2D DFB lasers \cite{zhou2013,turnbull2003,stehr2003}).  In figure \ref{fig:fourspecsbanddiagram}(h) we can see that there is also lasing on the third order bragg condition, as the frequency for the 3rd order condition has moved down into the gain window of Rh6G for a pitch of 400 nm. Polarization dependent measurements reveal that the lasing mode is a TE mode,  corresponding to what has been reported in literature for dielectric DFB lasers.\cite{turnbull2001,heliotis2004} 

\begin{figure*}
\centerline{\includegraphics[width=\widefigwidth]{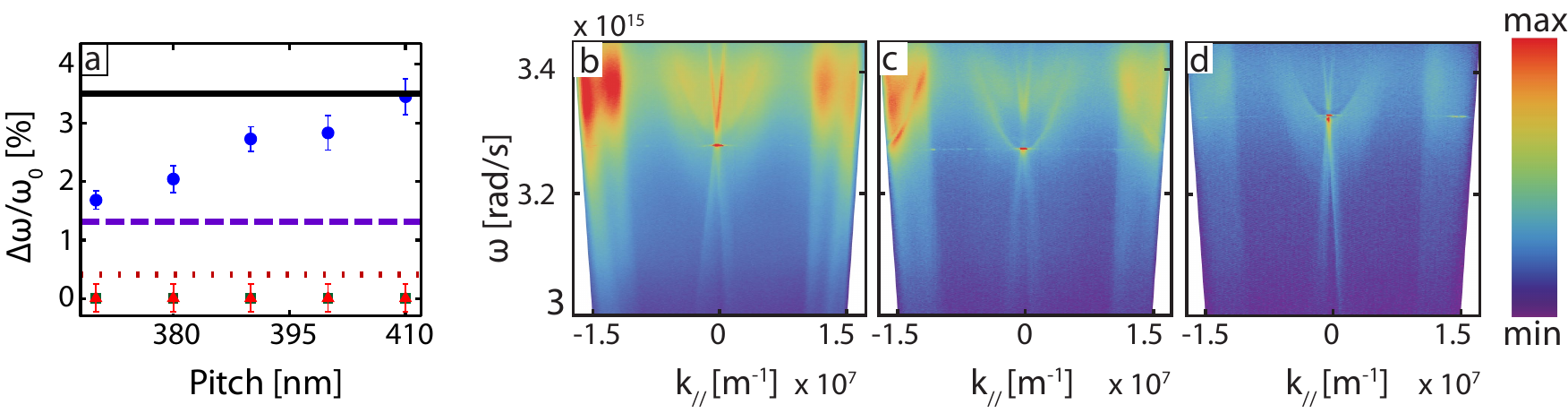}}
\caption{\label{fig:comparematerials} In a) the relative size of the stopgap is plotted for Ag (blue dots), Au (red triangles) and TiO$_2$ (green squares) together with literature values of van Beijnum \cite{beijnum2013} (black), Turnbull \cite{turnbull2003} (purple) and Noda \cite{noda} (red). In addition, fluorescence in the $\omega$, $k_y$  plane is plotted for Au (b), Ag (c) and TiO$_2$ (d) for a pitch of 370 nm.  The maximum values of the colorbar are 550, resp. 250 and 300 counts in panels (b-e), with colorbars starting at 0. The plots are made just above the lasing threshold, meaning pump powers differ from panel to panel. The pump powers are 944 nJ (Au), 99.5 nJ (Ag),   and 38 nJ (TiO$_2$), respectively. The noteably larger pump power for Au results in a much higher background fluorescence level as is visible for a frequency range centered at the the Rh6G emission peak in figure b. }
\end{figure*}
 Broadly speaking, it appears that the Ag particle array laser is close to a standard DFB laser in that it operates at the lower edge of the second diffraction stopgap at $\mathbf{k}_{||}=0$ (stop gap edges indicated by white ticks in panel \ref{fig:fourspecsbanddiagram}(f)). We now ask how the plasmonic nature of the Ag particles modify the DFB characteristics compared to a dielectric DFB laser \cite{lozano2013,murai2012}. %On one hand one might expect that the only difference would be a higher lasing threshold owing to the need to overcome Ohmic loss. On the other hand, one might expect that the much stronger scattering strength of plasmon particles implies a clearly distinct waveguide dispersion diagram, as indicated by extinction data on particle arrays reported by Lozano et al. \onlinecite{lozano2013} and Murai et al. \onlinecite{murai2012}.
 In order to probe this question we compare three systems, namely (1) the Ag particle arrays, (2) arrays of the same pitch of dielectric TiO$_2$ particles, and (3) Au particles that should show stronger absorption yet weaker scattering than the Ag particles. In figure \ref{fig:comparematerials} we compare the dispersion diagram for a DFB laser with silver scatterers with a DFB laser that uses TiO$_2$ and Au particles, for a particle pitch of 380 nm and a particle size of 150 nm and 100 nm, respectively. For TiO$_2$ we used larger particles of 150 nm diameter, as the TiO$_2$ particle arrays with particle sizes of 100 nm did not show lasing. This we assign to the weak scattering strength of 100 nm TiO$_2$ disks.
 Indeed, numerical analysis reported below of scattering cross sections show that TiO2 scatterers of the same volume have a scattering strength at least 10 times lower than metal particles.  
   Two clear differences are visible between figure \ref{fig:comparematerials}(c) and figure   \ref{fig:comparematerials}(d). For the TiO$_2$ sample, the parabolic band appears to be a single feature, as expected from figure \ref{fig:fourspecsbanddiagram}(c). However, strictly speaking the parabolic band is degenerate, originating from both the (1,0) and (-1,0) diffraction order. Remarkably, for the Ag particle array this degeneracy is distinctly split, pointing at the strong scattering strength of Ag particles. 
 Figure \ref{fig:comparematerials}(b)  shows the measured dispersion diagram for an Au array. Evidently, the bands are broad, at least as much as in the Ag case, but not clearly split, as in the TiO$_2$ case. This points at the higher loss, yet weak scattering strength at 590 nm, of Au particles compared to Ag. 
 
 For dielectric photonic crystals,  relative stop gap width  $\Delta \omega/\omega_0$  is frequently used as a dimensionless parameter to sort photonic crystals by their  photonic interaction strength.\cite{Vos96,Soukoulis2001bookp194,BenistyProgOptics49Ch4} In that case the relative stop gap width is  proportional to the ratio of scatterer polarizability to unit cell volume. In real space terms, the stop gap width provides a direct measure for the Bragg length (number of lattice planes required for 1/e diffraction efficiency), the crystal size needed to develop a significant LDOS suppression,  and the crystal size required to achieve   LDOS   enhancement of mode density at a  band edge of any significant magnitude and over significant bandwidth.\cite{Vos96,Soukoulis2001bookp194,BenistyProgOptics49Ch4}  On this basis, we  use the width of the stopgap to quantify differences in the dispersion diagram for different plasmonic laser systems. 
 Figure \ref{fig:comparematerials}(a) shows a plot of the relative size of this stopgap as a function of particle pitch for Ag, Au and TiO$_2$. The horizontal lines indicate relative stopgap values reported by van Beijnum \cite{beijnum2013} (black), Turnbull \cite{turnbull2003} (purple) and Noda \cite{noda} (red)for a plasmonic hole array laser, a non-plasmonic DFB laser, and a photonic crystal band edge laser in a 2D semiconductor membrane. The blue dots for Ag show that the relative bandgaps are large compared to reported values for dielectric systems and approach the value reported for the  plasmonic laser of van Beijnum \cite{beijnum2013}.  The red triangles and green squares represent relative bandgaps for TiO$_2$ and Au. For these two materials the stopgaps are smaller than the width of the band and therefore are essentially zero. To conclude, the Ag particle array DFB lasers are markedly different from the nonplasmonic lasing systems, and due to the strong scattering the diffractive coupling in the dispersion relation is as strong as in the plasmonic hole array laser. Ideally to verify if the correlation between stop gap width and sample geometry is exclusively with scattering strength (or `polarizability', as in the photonic crystal case)  one would need to independently vary physical particle volume at fixed optical volume (polarizability), or vice versa, in an otherwise fixed gain medium. Unfortunately, this will be difficult to realize: while our data clearly show that the nonplasmonic TiO$_2$ particles of equal physical volume are so weakly scattering as to give neither stop gap nor lasing, a larger optical volume at fixed physical size than for the silver particles can not be realized in the gain window of our dye. Somewhat larger optical volume could be reached with either Ag or Au particles by increasing their size,  however only at strongly redshifted resonance frequencies.

 Finally we note that in order to cross the lasing threshold, the Au samples typically require at least ten times higher pump fluence (0.066  mJ/cm$^2$ versus 1.107 mJ/cm$^2$ for the example in figure \ref{fig:comparematerials}). This finding is consistent with the much more advantageous Ohmic loss of Ag versus Au, which for Au nanodisks means a much lower scattering strength and a much lower albedo. Regarding the comparison between thresholds of the TiO$_2$ sample with the thresholds of Ag and Au samples, we have to note that a direct comparison is hampered by the fact that the much lower scattering strength of TiO$_2$ disks means that much larger particles were required than for the plasmonic samples to reach the lasing transition at all. Generally, most all-dielectric samples that actually lased (particle diameters above 150 nm) had lower thresholds than their plasmonic counterparts. This finding indicates that TiO$_2$ offers low loss, yet also a much weaker per-particle cross section contributing to feedback.

\section{Theory\label{section:theory}}
\subsection{Single particle scattering}{\label{section:comsol}
We use COMSOL to determine the extinction cross sections for single particles. In figure \ref{fig:extinction} we plot the extinction cross section for Ag, Au and TiO$_2$ for two different incidence conditions. In figure \ref{fig:extinction}(a) an $x$-polarized plane wave is incident along the $z$-direction (parallel to the symmetry axis of the particle disk). In figure \ref{fig:extinction}(b) an $x$-polarized plane wave is incident along the $y$-axis.  We used an index of $n=1.65$ for SU8 as the surrounding medium. For the permitivity we use a modified Drude model fitted to the optical constants of Johnson and Christy~\cite{Johnson1972}:
\begin{equation}
\epsilon_r=\epsilon_\infty -\frac{\omega_p^2}{\omega\cdot(\omega+i\gamma)}
\end{equation}
\noindent For Au we use $\epsilon_\infty = 9.54$, $\omega_p  = 1.35\cdot 10^{15}$ rad/s, and $\gamma= 1.25 \cdot 10^{14}$ rad/s and for Ag we use $\epsilon_\infty = 5.43+0.55i$, $\omega_p  = 1.39 \cdot 10^{16}$ rad/s, and $\gamma  = 8.21 \cdot 10^{13}$ rad/s, as reported in Ref.~\onlinecite{penninkhof2008}. 

\begin{figure}
\includegraphics[width=\columnwidth]{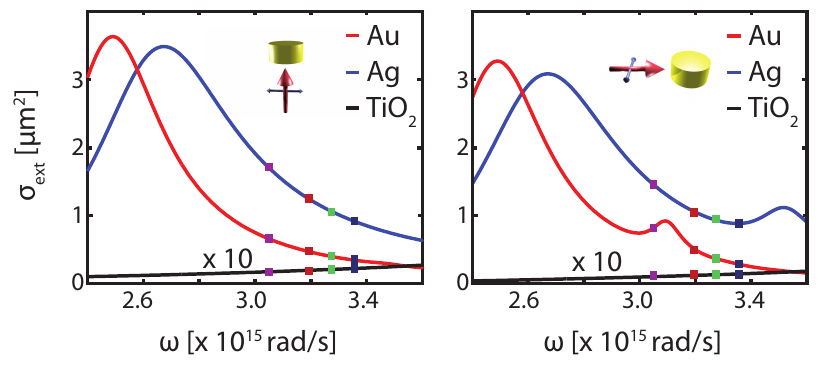}% Here is how to import EPS art

\caption{\label{fig:extinction}Extinction cross sections for Ag, Au and TiO$_2$ as a function of $\omega$ obtained using COMSOL. The z-axis is defined as the axis parallel to the symmetry axis of the particle disks. In plot a) the plane wave is incident along the z-axis and in plot b) the plane wave is incident from the side of the particle, with polarization in the plane of the particle. For clarity the curves for TiO$_2$ are scaled by a factor 10 as indicated.}
\end{figure}
Both Ag and Au exhibit a clear resonance which is completely absent for TiO$_2$. In addition,  one can see that the peak of Ag is blue shifted with respect to the resonance peak of Au. The lasing frequencies for the studied 4 particle pitches are indicated by the squares. For a plane wave along the y-direction we can distinguish 2 peaks, where the smallest peak (at higher frequencies) corresponds to the quadrupolar resonance. For incidence normal to the disks, as would be the case in transmission experiments that probe the sample under normal incidence, no quadrupole response is noticeable. However, for the distributed feedback in-plane scattering of the TE-polarized waveguide mode is important. This distributed feedback hence   might benefit from the quadrupole response for enhanced scattering and near fields.
\subsection{Band structure}{\label{section:rcwa}
Finally we have calculated the band structures of the plasmon gratings embedded in the waveguide structures as they would appear in extinction, using rigorous coupled wave analysis (RCWA) that is optimized for 2D periodic and stratified problems. In particular, we have used the freely available implementation ``S$^4$" by Liu and Fan~\cite{Liu2012} of the Fourier Modal Method developed by Li \cite{li1997,li1996}, that uses the appropriate factorization rules for high index contrast gratings. 
While convergence can be notoriously poor for metallic gratings, we found excellent convergence when using parallelogrammic truncation. We used a truncation to 361 plane waves. We set $n_{\mathrm{SU8}}=1.65$ and take the particle sizes and dielectric constants as in COMSOL. The index of the glass substrate is set to n=1.51, while we take as waveguide thickness 450 nm. Again, we use the Drude model to describe the permittivity of Ag and Au. We obtain extinction as a function of incidence angle, resulting in the extinction dispersion diagrams shown in figure \ref{fig:s4}.  While the Fourier modal method is a fully vectorial method that takes coupling  between Bloch harmonics of all polarizations into account, here we report specifically on extinction in the case of  s-polarized incidence, corresponding to coupling to TE-waveguide modes. This choice is motivated by our observation of the polarization of the lasing mode, and is commensurate with 2D DFB lasing in dielectric structures~\cite{turnbull2001,heliotis2004}.
Both the Ag and Au lattice  show a band of high extinction at $k_{||}=0$ close to $\omega  = 2.5 \cdot 10^{15}$ rad/s, where as expected the gold array is redshifted compared to the Ag array. This  extinction band corresponds to the single particle dipole resonance. Compared to figure \ref{fig:extinction}, figure \ref{fig:s4} shows single particle resonances slightly red shifted as in S$^4$ simulations infinite particle arrays are considered.  When the interparticle distance is comparable to the wavelength, longitudinal coupling between the individual dipole scatterers is known to cause the observed red shift \cite{sersic2009}. The higher extinction due to the single particle resonance couples to the waveguide mode \cite{murai2012}, leading to anticrossings at   $k_{||}$ along the straight lines corresponding to  the backfolded waveguide mode dispersion.  This anticrossing was observed experimentally  by Rodriguez et al.~\cite{rodriguez2012}. 
\begin{figure}
\includegraphics[width=\columnwidth]{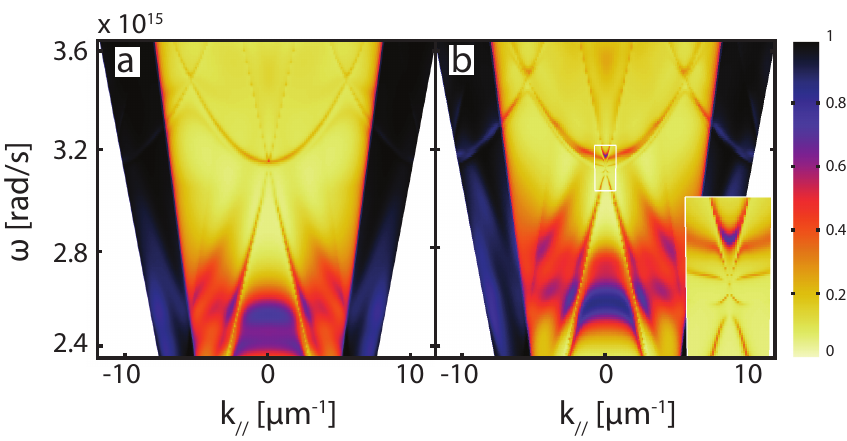}% Here is how to import EPS art
\caption{\label{fig:s4}Extinction as a function of k$_{||}$ and $\omega$ for gold (a) and silver (b). Evident for both diagrams are 1) the high extinction region for lower $\omega$, corresponding to the single particle resonance, and 2) the generic folded band diagram features (lines and parabola's).  For silver a clear stop gap is visible at the $\Gamma$ point ($\mathbf{k}_{||}=0$) that is not apparent in the diagram for gold. The inset shows a zoom in of this region, highlighting the intricate anticrossing with the two parabolic bands.}
\end{figure}
 At $3.2\cdot 10^{15}$ rad/s and above there are the expected parabola and straight lines from the folded free-photon dispersion discussed in the experimental results. In this part, we can see two clear differences between Au and Ag. First, the parabola for silver is broader than that of gold. 
Second, at the crossing point of the parabola with the straight lines a clear avoided crossing can be seen for silver, whereas for gold the $\Gamma$ point does not show a gap and corresponds with the folded free-photon dispersion. The complex avoided crossing that lifts the degeneracy between the two parabolas for silver can be seen more closely in the inset in figure \ref{fig:s4}(b).  While the single particle resonance is broad (Q=4.4), the hybrid modes resulting from coupling of waveguide mode and plasmon have Q> 150,  i.e.,  damping much less than the single particle radiative damping.

  Returning to a comparison with the measurements we conclude that the theory reproduces all the salient features. For the gold particle lattices,  the scattering strength per particle is low. Consequently in both theory and data, the stop gap width is small.  For the silver particle lattice, however, the scattering strength per particle is much higher and consequently in both theory and experiment a clear stop gap opens up at the second order diffraction, and the degenerate parabolas split, and broaden.  Based on the single particle response we surmise that the precise coupling strength that splits the bands at $3.2 \cdot 10^{15}$  rad/s  and above is dependent on the quadrupole response. Finally we note that we also calculated dispersion for TiO$_2$ particle lattices. As in experiment, the calculated dispersion (not shown) only shows narrow features that essentially coincide with the folded SU8 waveguide dispersion.

\section{Conclusion and Outlook\label{section:conclusion}}
Overall,  it is remarkable that despite their loss,  silver particle array lasers provide lasing characteristics reasonably competitive with dielectric DFB systems, commensurate with the established notion that the extended diffractive modes that plasmon arrays support well off their individual resonance frequencies are promising for balancing loss and light matter interaction strength. Furthermore, our results and experimental methods open up many questions for further study. To start, it would be very interesting to sweep the diffraction condition and gain window onto the plasmon resonance frequency. This allows to continuously trace how lasing occurs along the transition from a weakly coupled plasmon-waveguide hybrid system to a purely plasmonic mode.  Second, if one could probe lasing in a given system for gain media of different quantum efficiencies, one could further clarify the role of high Purcell factors near metal particles in lasing plasmonic lattices. In our work, we estimate that less than 1\% of emitters is within 20~nm of  metal, i.e., is in a position where Purcell enhancement might occur. Since we use an intrinsically already highly efficient, bright, emitter,  there is no measurable Purcell enhancement, and in fact those dyes that experience rate enhancement likely are rendered \emph{less} efficient contributors to the lasing process due to quenching. This should be contrasted to the work of Suh et al. and Zhou et al. \cite{suh2012,zhou2013}.  Third, by significantly reducing the pitch, one enters the `lasing spaser' regime proposed by Zheludev et al. Zheludev et al.\cite{zheludev2008}  proposed that when plasmon resonators with gain are arrayed with pitch much smaller than the wavelength, lasing will not occur on a diffraction condition, yet coherence will be established to give lasing emission normal to the lattice plane. Generally we expect that if a dense metasurface would lase, it would do so on the lowest loss mode in the wave vector diagram. Hence this regime directly necessitates a deep study of the dispersion relation of collective modes in metamaterials. Finally, an obvious extension of our work is to study aperiodic systems \cite{vardeny2013,gopinath2008, lubin2013}. Previous studies have shown that  aperiodic and quasiperiodic systems, as an intermediate state between order and disorder, have modes that are neither Bloch states as in a periodic lattice, nor exponentially localized states as in a random array, but exhibit 'critical modes' that show strong spatial fluctuations in field amplitude \cite{macia2012,vardeny2013}. It would be interesting to study these critical modes in the context of lasing.

\acknowledgments
We thank Marko Kamp, Henk-Jan Boluijt, Marco Seynen en Jan Zomerdijk for help with the design of the optical set up,  the acquisition software, and the pulse integrator for power caliration.  Furthermore we are indebted to Felipe Bernal Arango for valuable insight in the calculations,  Clara Osorio and Erik Garnett for carefully checking the manuscript, and Said Rahimzadeh-Kalaleh Rodriguez and Jaime Gomez Rivas for discussion about luminescence in periodic particle systems.
This work is part of the research  program of the ``Foundation for Fundamental Research on Matter (FOM)'', which was financially supported by `The Netherlands Organization for Scientific Research (NWO)'. AFK gratefully acknowledges an NWO-Vidi grant for financial support. This work was furthermore supported by NanoNextNL, a micro and nanotechnology consortium of the Government of the Netherlands and 130 partners.

\bibliography{Schokker}% Produces the bibliography via BibTeX.

\end{document}